\documentclass[twocolumn,prb,showpacs,preprintnumbers,amsmath,amssymb,floatfix]{revtex4}

\usepackage{epsfig,psfrag}
\usepackage{dcolumn}
\usepackage{bm}
\usepackage{graphicx}
\usepackage{color}

\begin{document}

\title{Phonon-phonon interactions and phonon damping in carbon nanotubes}
\author{Alessandro De Martino,$^1$ Reinhold Egger,$^2$ and Alexander
O. Gogolin$^3$}
\affiliation{
$^1$ Institut f\"ur Theoretische Physik, Universit\"at zu K\"oln, Z\"ulpicher
Stra{\ss}e~77, D-50937 K\"oln, Germany\\
$^2$ Institut f\"ur Theoretische Physik, Heinrich-Heine-Universit\"at,
D-40225  D\"usseldorf, Germany\\
$^3$ Department of Mathematics, Imperial College London, 180 Queen's Gate,
London SW7 2AZ, UK }

\date{\today}

\begin{abstract}
We formulate and study the effective low-energy quantum theory of 
interacting long-wavelength acoustic phonons in carbon nanotubes 
within the framework of continuum elasticity theory. 
A general and analytical derivation of all three- and four-phonon processes 
is provided, and the relevant coupling constants are determined in terms of few 
elastic coefficients.
Due to the low dimensionality and the parabolic dispersion, 
the finite-temperature density of noninteracting flexural phonons
diverges, and a nonperturbative approach to their interactions
is necessary.  Within a mean-field description, we find that a dynamical 
gap opens. In practice, this gap is thermally smeared, but still has 
important consequences.  Using our theory, we compute
the decay rates of acoustic phonons due to
phonon-phonon and electron-phonon  interactions, 
implying upper bounds for their quality factor.  
\end{abstract}


\maketitle

\section{Introduction}\label{sec1}

Even after more than a decade of very intense research efforts,\cite{charlier}
the unique electronic and mechanical properties of carbon nanotubes (CNTs)
continue to attract considerable interest. A major driving force for this
interest comes from the prominent role played by phonons in CNTs.
Phonons are crucial when interpreting experimental 
data for resonant Raman or photoluminescence excitation 
spectra,\cite{dresselhaus,rao} and for the understanding
of electrical\cite{roche} and thermal\cite{kim} transport in CNTs. 
Moreover, phonons are responsible for interesting 
nanoelectromechanical effects in suspended 
CNTs,\cite{sapmaz,roy,sazonova,huang,zant} and 
they lead to quantum size effects in the specific heat.\cite{hone}
The real-time nonlinear dynamics of a CNT phonon mode has also been
monitored experimentally by femtosecond pump-probe techniques (coherent phonon 
spectroscopy).\cite{gambetta,sanders} 

Recent experiments have shown
that mechanical oscillations of suspended carbon nanotubes can be 
excited by a cantilever and detected by scanning force 
microscopy.\cite{babic,bachtold} Such experiments yield both the 
frequency $\omega$ and the quality factor $Q=\omega/\Gamma$ 
(with decay rate $\Gamma$) of the respective phonon mode.  
The so far observed\cite{bachtold} values,  $Q \alt 10^4$, imply significant 
decay rates even at rather low temperatures, and require to 
identify the relevant decay channels for phonons in individual CNTs.
Our paper is primarily devoted to understanding the importance
of phonon-phonon interactions in such decay processes.  
The quality factor can also be extracted from Raman spectroscopy\cite{rao} and
from nanoelectromechanical measurements, using phonon-assisted Coulomb blockade 
spectroscopy\cite{roy} or capacitive detection of mechanical 
oscillations.\cite{sazonova} In principle,
coherent phonon spectroscopy also allows to
access damping rates of phonon modes, and hence their quality factors.
Very recently, the possibility of cooling a vibrating carbon nanotube
to its phononic ground state has also been discussed.\cite{bachtold09}

The recent experimental progress described above highlights the
need for a reliable theory of phonon-phonon (ph-ph) interactions in CNTs.
On the theoretical side, many authors have analyzed
the noninteracting problem, i.e. the harmonic (or linear) theory,
which allows to derive explicit theoretical results for the thermal 
conductance\cite{yamamoto,mingo1} and for the specific heat.\cite{zhang}
Motivated by the observation that molecular dynamics calculations
seem to be in good agreement with thin-shell model predictions,\cite{yakobson}
several theoretical works\cite{thinshell,mahan1,goupalov,chico,walgraef} 
have adapted thin-shell hollow cylinder models\cite{love,yamaki} to the 
calculation of phonon spectra. 
However, the thin-shell approach leaves open the question of how
to actually choose the width of the carbon sheet. 
A popular and more microscopic approach is to instead start from 
force-constant models,\cite{saito,mahan2,jiang} taking 
into account up to fourth-nearest-neighbor couplings in the most
advanced formulations.\cite{white,zimmermann}  These calculations  
predict four acoustic phonon branches (with $\omega(k\to 0)=0$), namely a
longitudinal stretch mode,  a twist mode, and two degenerate 
flexural modes (see Sec.~\ref{secIIC} for details).  
The resulting phonon spectra are in very good agreement with a much 
simpler calculation based on continuum elasticity 
theory,\cite{suzuura,ademarti} 
building on the known elastic isotropy of the honeycomb lattice.\cite{landau} 
The elastic approach will be employed in our study as well. 
Ref. \onlinecite{crespi} 
provides a general discussion of the accuracy of
elastic continuum theories for phonons in CNTs. 
%including benchmarks against first-principle calculations. 
For very small CNT radius $R$, however,
hybridization effects involving carbon $\sigma$ orbitals 
lead to qualitative changes, elastic continuum theories (at least
in the form below) may break down, and 
first-principle calculations become necessary.\cite{kresse,barnett}

In contrast, the problem of ph-ph interactions in CNTs 
is much more difficult and has been treated in only a few works,
although phonon anharmonicities are important for several physical 
observables,\cite{ashcroft,maradudin,cowley} e.g. 
for thermal expansion (which is 
a controversial issue in CNT theory\cite{thermal2}), to explain
the stability of low-dimensional materials (which would be unstable
in harmonic approximation due to the Mermin-Wagner theorem), in order
to establish a finite thermal conductivity, or to provide a finite
lifetime for phonons.  The latter issue is particularly relevant in our context, but  
aside from a numerical high-temperature study\cite{hepplestone} which 
ignored the (lowest-lying) flexural phonons, 
to the best of our knowledge ph-ph interactions in CNTs have only 
been studied by Mingo and Broido.\cite{mingo1,mingo2} 
Their work considered three-phonon processes and their effects
within a Boltzmann transport equation.  The main conclusion
of Refs.~\onlinecite{mingo1}, \onlinecite{mingo2} was that anharmonic effects
are generally weak but important in establishing upper bounds
for the thermal conductance. In addition, they computed
the lengthscale up to which phonons show ballistic motion. 
Where applicable, our results below are in accordance with theirs, 
but four-phonon processes (which govern the decay of flexural 
phonons) have not been studied so far.

We shall consider two important mechanisms for the decay of 
long-wavelength acoustic phonons in single-wall CNTs, 
namely electron-phonon (e-ph) and ph-ph scattering.  
We show that the dominant e-ph coupling terms (resulting 
from the deformation potential contribution) 
do not allow for phonon decay due to 
kinematic restrictions, and thus an intrinsic upper bound for the 
temperature-dependent quality factor of the various modes can be 
derived from ph-ph interactions alone.  These upper bounds are given below. 
The problem of phonon decay has in fact a rather long history.
Early work on the decay of an optical phonon into
two acoustic phonons via anharmonicities\cite{orbach,klemens} 
proposed a scheme for nonlinear phonon generation.
Phonon decay via ph-ph interaction is also important 
for the understanding of neutron scattering data\cite{maradudin} and for
the collective excitations in
liquid helium.\cite{maris} Such effects have even been considered
in a proposal for a phonon-based detector of dark matter.\cite{tamura}
General kinematic restrictions often prevent the decay
of phonon modes. Lax \textit{et al.} have shown\cite{lax}
that a given acoustic phonon  cannot decay into other
modes with higher velocity at any order in the anharmonicity.  
For the lowest-lying acoustic phonon mode, one then expects 
\textit{anomalously long lifetimes}, while the 
higher acoustic modes typically decay --- in three-dimensional (3D)
isotropic media with rate $\Gamma\propto |{\bf  p}|^5$ for 
phonon momentum ${\bf p}$.\cite{landau,tamura2}
Such questions are particularly interesting in the CNT context,
where a degenerate pair of flexural modes has the lowest energy, 
and the low dimensionality and the quadratic dispersion relation
of the flexural mode may give rise to unconventional behavior.

Before describing the organization of the paper,
we pause for some guidance for focused readers. 
Experimentally minded readers can find our central predictions for the 
decay rate (and hence the quality factor) of the low-energy
phonon modes in Eqs.~(\ref{gamml}), (\ref{gammt}) and (\ref{finalrate}).
The dependence of the resulting $Q$ factors on the CNT radius $R$
is shown in Fig.~\ref{fig2}.  Those interested in the 
main new theoretical results will find them in  Eqs.~(\ref{3phonon}) 
and (\ref{4phonon}), 
where the complete low-energy Hamiltonian for 
interacting acoustic phonons in single-wall CNTs is given, 
with the modified flexural dispersion relation (\ref{flexnew}). This
modification takes into account the instability of a harmonic
theory implied by the Mermin-Wagner theorem, and includes 
interaction effects in a nonperturbative manner. The calculation
of the decay rates is then possible in a perturbative manner, and
leads to the results quoted above. 

Let us conclude this Introduction with the organization of the paper.
In this work, based on the elastic continuum description,
we formulate a complete and analytical theory of interacting
long-wavelength acoustic phonons in single-wall CNTs.
In Section \ref{sec2} we show that the simplicity of the elastic approach 
allows us to go beyond the harmonic approximation 
(which is briefly reviewed in Sec.~\ref{secIIC}), 
and thereby provides a complete theory of all possible 
three- and four-phonon scattering processes, described
in detail in Sec.~\ref{sec3}. 
The theory is then applied in Sec.~\ref{sec4} to the calculation
of phonon decay rates. We thereby infer intrinsic upper bounds for
the quality factor of the relevant acoustic modes.
We comment on effects of e-ph interactions on the 
quality factor in Sec.~\ref{sec5}, and end the paper with a discussion and an 
outlook in Sec.~\ref{sec6}. 
Calculational details have been relegated to two appendices.
Finally, we note that while some of our results are also relevant
to 2D graphene monolayers,\cite{bonini,mariani,kinaret} 
for the sake of clarity, we restrict ourselves to the CNT case
throughout the paper. We sometimes set $\hbar=1$ in intermediate steps.

\section{Nonlinear strain tensor and elastic theory}\label{sec2}

In this section we shall develop the low-energy theory of interacting 
long-wavelength acoustic phonons in CNTs. To be specific, we first 
discuss \textit{semiconducting single-wall CNTs}, where e-ph 
scattering processes can safely be ignored. 

\subsection{Strain tensor in cylindrical geometry}

We start from a continuum description, where long-wavelength
phonons are encoded in the  three-dimensional \textit{displacement 
field}, $\bf u$, with local-frame components $u_{n=x,y,z}$ (see below).
The surface of an undeformed cylinder, representing 
the CNT with radius $R$, is parametrized as 
\begin{equation}\label{rdef}
{\bf R}({\bf r}) = R {\bf e}_z (x) + y {\bf e}_y, \quad {\bf r}=(x,y) .
\end{equation}
We use cylindrical coordinates with $x/R$ (where $0\leq x<2\pi R$)
denoting the angular variable. The corresponding 
local-frame unit vector is ${\bf e}_x (x)$, while
${\bf e}_y$ points along the cylinder axis and ${\bf e}_z(x)$ is
perpendicular to the cylinder surface, i.e. $z$ corresponds to the
radial coordinate.  Note that $R\partial_x {\bf e}_z= {\bf e}_x$ and 
$R\partial_x {\bf e}_x=-{\bf e}_z$, and ${\bf R}$ depends only 
on the coordinates ${\bf r}=(x,y)$ parametrizing the cylinder surface.
Our convention for the coordinates follows the notation
of Ref.~\onlinecite{suzuura}, which is convenient because it
connects the problem on the cylinder (CNT) to the one on the plane (graphene).

The surface of the deformed cylinder is then parametrized in terms
of the displacement field as
\begin{equation}\label{xdef}
{\bf x}({\bf r})={\bf R} ({\bf r}) +  {\bf u}({\bf r}) = {\bf R} ({\bf r}) 
+\sum_{n=x,y,z} u_n({\bf r}) {\bf e}_n(x) .
\end{equation}
Equations (\ref{rdef}) and (\ref{xdef}) imply the relation
\begin{eqnarray} \label{dx}
d{\bf x}&=&   \left[du_x+ \left(1+\frac{u_z}{R }\right) dx\right] {\bf e}_x(x) 
\\ \nonumber & +& [du_y+dy] {\bf e}_y + \left[du_z-\frac{u_x}{R}dx\right] 
{\bf e}_z(x) ,
\end{eqnarray}
where contributions come both from the variation of the displacement field and
from the change in the local frame.  Given the displacement field, the  
symmetric \textit{strain tensor} $u_{ij}({\bf r})$, with $i,j=x,y$,
 can be obtained
from the defining relation\cite{landau} 
\begin{equation}\label{defrel}
u_{xx} dx^2 + u_{yy} dy^2 +2 u_{xy} dxdy=
\frac12 (d{\bf x}^2-d{\bf R}^2).
\end{equation}
Employing Eq.~(\ref{dx}), after some algebra, the strain tensor follows.
It is composed of a linear and a nonlinear part in the displacement field,
$u = u^{\rm lin} + u^{\rm nlin}$, \begin{eqnarray} \label{ulin}
2u^{\rm lin}_{ij} &=&  D_i u_j + D_j u_i, \\ 
\label{unlin} 2 u^{\rm nlin}_{ij} &=&\sum_{n=x,y,z} (D_i u_n) (D_j u_n),
\end{eqnarray}
where we use covariant derivatives, 
\begin{equation} \label{cov1}
D_x u_x = \frac{\partial u_x}{\partial x}  + \frac{u_z}{R},
\quad D_x u_z = \frac{\partial u_z}{\partial x} -  \frac{u_x}{R},
\end{equation}
while $D_y u_n=\frac{\partial}{\partial y} u_n$ and 
$D_x u_y=\frac{\partial}{\partial x}u_y$.

One easily verifies that the strain tensor $u_{ij}$ respects fundamental symmetries. 
In particular,  $u_{ij}=0$ for arbitrary rigid translations or 
rotations of the whole cylinder.  For translations, 
both the linear and the nonlinear part of the
strain tensor vanish separately, but this is not the case
for rotations. While $u^{\rm lin}_{ij}=0$ under infinitesimal rotations, 
the full nonlinear strain tensor must be kept in order to 
correctly account for $u_{ij}=0$ under \textit{finite} rotations.

\subsection{Elastic energy density}

The Hamiltonian density is given by the sum of the kinetic and the
elastic energy density,
\begin{equation}
{\cal H} = \frac{1}{2\rho_M}\sum_{n} p_n^2 + {\cal U}[u],
\end{equation}
where $\rho_M=3.80\times 10^{-7}$~kg$/$m$^2$ is the mass density of 
graphene, and $p_n$ is the canonically conjugate momentum to 
$u_n$. The theory is quantized via the standard commutation 
relations [with ${\bf r}=(x,y)$ and $n, n'=x,y,z$],
\begin{equation}\label{comm}
[ p_n ({\bf r}), u_{n'} ({\bf r}')]_- = 
-i\hbar \delta_{nn'} \delta({\bf r}-{\bf r}').
\end{equation}
Armed with the nonlinear strain tensor, progress is now possible by invoking 
symmetry considerations and the usual assumption of a space-time local 
elastic energy density ${\cal U}= {\cal U} [u({\bf r},t)]$ depending only
on the strain tensor.
The elastic energy density is then expanded in the strain tensor up to fourth order,
${\cal U}[u]={\cal U}_2+ {\cal U}_3+ {\cal U}_4$.
This expansion will fully account for all elementary
ph-ph scattering processes involving at most four phonons.
In order to give explicit expressions, we will exclude the
case of ultrathin CNTs, which is difficult to model with 
an elastic continuum approach.
For extremely small radius, $\sigma-\pi$ orbital hybridization effects 
due to the curvature of the cylinder can lead to dramatic
effects and, in particular, may change the honeycomb lattice
structure.\cite{barnett}  In practice, this means 
that we require $R\agt 4$\AA.
In that case, curvature effects generally scale as $1/R^2$, and 
as outlined in Appendix \ref{appA}, their inclusion 
is possible on phenomenological grounds within our nonlinear elasticity
theory.  Ignoring curvature effects for the moment, a straightforward 
connection to the corresponding planar problem of graphene can be established.  
Since graphene's honeycomb lattice is \textit{isotropic} 
with respect to elastic properties,\cite{landau} ${\cal U}[u]$ can only depend on 
invariants of the strain tensor under the symmetry group $O(2)$. 
Independent invariants can then be formed using the trace of the 
strain tensor, ${\rm Tr} u$, and its determinant,  ${\rm det}\ u= 
[({\rm Tr} u)^2  - {\rm Tr}  u^2]/2 $.

Starting with ${\cal U}_2[u]$, quadratic in the strain tensor,  one arrives
at the familiar expression\cite{landau}
\begin{equation}\label{u2}
{\cal U}_2[u]=\frac{\lambda}{2}\left( {\rm Tr} u \right)^2 +\mu \ {\rm Tr} u^2,
\end{equation}
where $\lambda$ and $\mu$ are \textit{Lam\'e coefficients}.
Their value in graphene is estimated to be 
(see, e.g., Refs.~\onlinecite{suzuura,rubio})
\begin{equation}\label{lame}
\frac{K}{\rho_M} \simeq 2.90\times 10^8 \frac{{\rm m}^2}{{\rm s}^2}, \quad
\frac{\mu}{\rho_M} \simeq 1.51\times 10^8~\frac{{\rm m}^2}{{\rm s}^2},
\end{equation}
with the bulk modulus $K=\mu+\lambda$. The 2D Poisson ratio then corresponds to 
\begin{equation}\label{poisson}
\nu= \frac{K-\mu}{K+\mu} \simeq 0.31,
\end{equation}
in agreement with that computed using an empirical 
force-constant model.\cite{lu}
Note that ${\cal U}_2[u]$ is already \textit{nonlinear} in the
displacement field due to the nonlinearity (\ref{unlin}) of the strain tensor. 
We shall refer to such nonlinearities, resulting already from ${\cal U}_2$, 
as \textit{geometric}. Geometric nonlinearities do not involve new material 
parameters beyond the Lam\'e coefficients.  

Apart from geometric nonlinearities, there are also anharmonic contributions 
(${\cal U}_3$ and ${\cal U}_4$) due to higher-order terms in the 
expansion of the elastic energy density in the strain tensor.
In cubic order, one can build three invariants from
the strain tensor, namely ${\rm Tr} u^3, {\rm Tr} u^2 \ {\rm Tr} u$, and
$({\rm Tr} u)^3$. However, these invariants are not independent, since 
$2 \ {\rm Tr} u^3 = 3 \ {\rm Tr}u \ {\rm Tr}u^2 - ({\rm Tr}u)^3$. 
Hence there are just two new  anharmonic couplings
 in third order, denoted as $\xi_1$ and
$\xi_2$, leading to 
\begin{equation}\label{u3}
{\cal U}_3 [u] = \xi_1 ({\rm Tr} u)^3 + \xi_2 \ {\rm Tr} u^2 \ {\rm Tr} u.
\end{equation}
Taking $u=u^{\rm lin}$, this produces three-phonon interaction processes
from anharmonic terms in the elastic energy density, on top of the 
geometric nonlinearities.  Note that the nonlinear part $u^{\rm nlin}$ 
then causes four-phonon processes (or higher orders) from Eq.~(\ref{u3}).
Finally, in quartic order there are five invariants,
\[
{\rm Tr} u^4, \quad {\rm Tr} u^3 \ {\rm Tr}  u, \quad ({\rm Tr}  u^2)^2, 
 \quad {\rm Tr}u^2  \ ({\rm Tr}  u )^2, \quad ( {\rm Tr}  u )^4 .
\]
However, because of the identities
\begin{eqnarray*}
2 {\rm Tr} u^4  &=& -( {\rm Tr} u )^4 + ( {\rm Tr} u^2 )^2 + 2 ({\rm Tr} u )^2 \
 {\rm Tr} u^2 , \\ {\rm Tr} u \ {\rm Tr} u^3 &=& {\rm Tr} u^4 - \frac{1}{2} \left(
( {\rm Tr} u^2 )^2 -  ( {\rm Tr} u)^2  \ {\rm Tr} u^2 \right) ,
\end{eqnarray*}
only three out of the five invariants are independent.  With 
fourth-order anharmonic couplings $(\kappa_1,\kappa_2,\kappa_3)$, we can thus write
\begin{equation}\label{u4}
{\cal U}_4 [u] = 
\kappa_1 ({\rm Tr} u)^4 + \kappa_2 ({\rm Tr} u)^2 \ {\rm Tr} u^2
+ \kappa_3 ({\rm Tr} u^2)^2.
\end{equation}
As we show below, cf.~Eqs.~(\ref{gamml}), (\ref{gammt}) and 
(\ref{finalrate}), the dominant decay processes for acoustic
phonons are governed by the geometric nonlinearities 
alone, and no parameter estimates for the anharmonic 
couplings ($\kappa_{1,2,3}$ and $\xi_{1,2}$) are necessary for
the calculation of these decay rates. This remarkable result 
could not have been anticipated without explicit computation of all
contributions.

The final step is to insert $u=u^{\rm lin}+u^{\rm nlin}$ in ${\cal U}[u]$,
and thereby to separate the harmonic theory (noninteracting phonons, $H_0$)
from interactions (three-phonon, $H^{(3)}$, and four-phonon, $H^{(4)}$, processes),
where $H=H_0+H^{(3)}+H^{(4)}$.
We do not take into account higher-order ph-ph scattering processes
beyond the fourth order.  
Collecting terms, the harmonic theory corresponds to the Hamiltonian density
\begin{equation}\label{h0}
 {\cal H}_0 = \frac{1}{2\rho_M} \sum_n p_n^2 + \frac{\lambda}{2} \left( 
{\rm Tr}u^{\rm lin} \right)^2 +\mu \ {\rm Tr} \left[ (u^{\rm lin})^2 \right].
\end{equation}
All possible three-phonon processes are encoded in
\begin{widetext}
\begin{equation}\label{h3}
{\cal H}^{(3)} = \lambda \ {\rm Tr} u^{\rm nlin} \ {\rm Tr} 
u^{\rm lin}+2\mu \ {\rm Tr}( u^{\rm lin} u^{\rm nlin}) 
 + \xi_1  \left({\rm Tr} u^{\rm lin}\right)^3
+ \xi_2 \ {\rm Tr} u^{\rm lin}\ {\rm Tr} \left[ (u^{\rm lin})^2 \right ],
\end{equation}
while all four-phonon processes are contained in
\begin{eqnarray}
\nonumber
{\cal H}^{(4)} &=& \frac{\lambda}{2}
 \left( {\rm Tr} u^{\rm nlin} \right)^2 +\mu \ {\rm Tr}\left[ (
u^{\rm nlin})^2 \right] +
3\xi_1 {\rm Tr} u^{\rm nlin} \left( {\rm Tr} u^{\rm lin} \right)^2 
+ \xi_2 {\rm Tr} u^{\rm lin} \left[ 2 {\rm Tr}(u^{\rm lin} u^{\rm nlin})
+ {\rm Tr} u^{\rm nlin} \ {\rm Tr} u^{\rm lin} \right] \\ 
\label{h4}
&+& \kappa_1 ({\rm Tr} u^{\rm lin})^4  +
\kappa_2 ({\rm Tr} u^{\rm lin})^2 \ {\rm Tr}\left[ (u^{\rm lin})^2 \right]
+ \kappa_3 \left({\rm Tr} \left[ (u^{\rm lin})^2 \right]\right)^2.
\end{eqnarray}
\end{widetext}
While this may seem like a rather complicated theory, we shall
see below that the geometric nonlinearities (i.e. the terms corresponding to
the Lam\'e coefficients $\lambda$ and $\mu$) already generate the most
relevant structures. 

\subsection{Harmonic theory}
\label{secIIC}

Let us first diagonalize the noninteracting Hamiltonian $H_0$, see Eq.~(\ref{h0}),
and thereby determine the phonon spectrum.  Although the results of this 
subsection have essentially been obtained before,\cite{suzuura,ademarti} we 
repeat the main steps in order to keep the paper self-contained. 
First, we perform a Fourier transformation of
the displacement field $u_n({\bf r})$, introducing the momentum $\hbar k$ along 
the $y$-axis and the integer angular momentum quantum number $\ell$,
\[
u_n({\bf r}) = \frac{1}{\sqrt{2\pi R}} 
\sum_{k,\ell}e^{i\ell x/R+iky} u_{n}( k, \ell), 
\]
where $u^\dagger_{n}(k,\ell)=u^{}_{n}(-k,-\ell)$ and $\sum_k \equiv 
\int_{-\infty}^\infty \frac{dk}{2\pi}$, and an analogous
transformation for $p_n$. The commutation relations (\ref{comm})
then read 
\[
[p_{n}(k,\ell), u_{n'}(-k',-\ell')]_-=-2\pi i\hbar \delta_{nn'}
 \delta_{\ell\ell'} \delta(k-k').
\]
Some algebra yields $H_0$ in the form
\begin{eqnarray}
H_0 = \int dx dy \, {\cal H}_0 &=& \frac{1}{2\rho_M} \sum_{  n k \ell } p^\dagger_{ n}( k, \ell)
p^{}_{n}( k,\ell ) \\ 
\nonumber &+& \frac12 \sum_{nn',k\ell} u^\dagger_{n}(k,\ell)
 \Lambda_{nn'}(k,\ell) u^{}_{n'}(k,\ell) ,
\end{eqnarray}
where the elastic matrix $\mathbf{\Lambda}(k,\ell)=(\Lambda_{nn'})(k,\ell)$ is
given by
\begin{equation}\label{elmat}
\mathbf{\Lambda}=\left( \begin{array}{lcr} \frac{\ell^2 (K+\mu)}{R^2}+
\mu k^2 & \frac{k\ell K}{R} & -\frac{i\ell (K+\mu)}{R^2}\\
 \frac{k\ell K}{R} & \frac{\ell^2 \mu}{R^2}+(K+\mu)k^2 & -\frac{ik(K-\mu)}{R}\\
 \frac{i\ell(K+\mu)}{R^2} & \frac{ik(K-\mu)}{R} & \frac{K+\mu}{R^2}
 \end{array} \right ) .
\end{equation}
This $3\times 3$ matrix is obviously Hermitian and obeys the time-reversal symmetry
relation\cite{landau} $\mathbf{\Lambda}^{}(-k,-\ell) 
= \mathbf{\Lambda}^{*}(k,\ell)$, where the star denotes complex conjugation.
Note that the chirality of the CNT does not affect
the elastic matrix (and hence the dispersion relation) within the 
low-energy theory. However, the situation is different for high-energy optical
phonons or when taking e-ph interactions into account.\cite{crespi}

The normal-mode frequencies $\omega_{J}(k,\ell)$ with corresponding polarization
unit vectors ${\bf e}^{}_{J}(k,\ell)$ (the index $J$ labels the normal modes) 
then follow from diagonalizing the elastic matrix,
\begin{equation}\label{eigen}
\mathbf{\Lambda}^{}(k,\ell) {\bf e}^{}_{J}(k,\ell) = 
\rho_M \omega_{J}^2(k,\ell)  {\bf e}^{}_{J} (k,\ell).
\end{equation}
The above symmetries of the elastic matrix imply $\omega_{J}(-k,-\ell)=
\omega_{J}(k,\ell)$ and $[{\bf e}^{}_{J}(k,\ell) ]^* = {\bf e}^{}_{J}(-k,-\ell)$.
Moreover, polarization vectors for given $k$ and $\ell$ are orthonormal,
${\bf e}^*_{J}(k,\ell) \cdot {\bf e}^{}_{J'}(k,\ell)=\delta_{JJ'}.$
Expanding the displacement field in terms of the polarization 
vectors and introducing boson creation, $a^\dagger_{J}(k,\ell)$, and annihilation,
$a^{}_J(k,\ell)$, operators,
\begin{equation}\label{algebra}
 [ a^{}_J(k,\ell), a^\dagger_{J'}(k',\ell') ]_-  = 
2\pi\delta(k-k') \delta_{JJ'}\delta_{\ell\ell'} ,
\end{equation}
we arrive at the quantized noninteracting phonon Hamiltonian,
\begin{equation}\label{nonint}
H_0 = \sum_{Jk\ell} \hbar \omega_{J}(k,\ell) \left( a^\dagger_{J}(k,\ell)
a^{}_J(k,\ell)+\frac12 \right).
\end{equation}
The displacement field components are then 
\begin{equation}\label{displ}
u_n({\bf r}) = \frac{1}{\sqrt{2\pi R}} \sum_{Jk\ell} 
e^{i\ell x/R + iky} [{\bf e}^{}_J(k,\ell)\cdot {\bf e}_n] u_{J}^{}(k,\ell),
\end{equation}
with the \textit{normal-mode components}, 
expressed in terms of the boson operators,
\begin{equation} \label{unewdef}
u_{J}^{}(k,\ell) = \sqrt{ \frac{\hbar}{2\rho_M \omega_{J}(k,\ell)} } 
\left(a^{}_J(k,\ell)+ a^{\dagger}_{J}(-k,-\ell)\right).
\end{equation}
We next summarize the solutions of the eigenvalue problem (\ref{eigen}).
We are interested in the long-wavelength ($|k|R\ll 1$) phonon
modes, in particular those with $\omega_{J}(k\to 0,\ell) =0$.

In the $\ell=0$ sector, there are three eigenmodes, namely $J=T$ (twist mode),
$J=L$ (longitudinal stretch mode), and $J=B$ (breathing mode).  
For the twist mode, we find for arbitrary $k$ the result
\begin{eqnarray}\label{twist}
\omega_{T} (k) &=& v_T |k| ,  \quad v_T=\sqrt{\frac{\mu}{\rho_M}}, \\ \nonumber   
{\bf e}_T(k) &=& {\bf e}_x = \left( \begin{array}{c} 1 \\ 0 \\ 0 
\end{array} \right),
\end{eqnarray}
where $v_T=1.23\times 10^4\,$m/s.
Note that  ${\bf e}_x$ points along the circumferential direction. 
For the longitudinal stretch mode, we obtain
\begin{eqnarray}\label{long}
\omega_L(k) &=& v_L |k| + {\cal O}(k^2),\quad v_L=
\sqrt{\frac{4K\mu}{\rho_M(K+\mu)}}, \\
\nonumber {\bf e}_L(k) &=&  
\left(  \begin{array}{c}
0 \\1 \\  - i\nu k R \end{array}\right)  + {\cal O}(k^2),
\end{eqnarray}
where $\nu$ is given in Eq.~(\ref{poisson}) and $v_L=1.99\times 10^4\,$m/s.
To lowest order in $|k|R$, ${\bf e}_L(k)$ points along the CNT axis ${\bf e}_y$,
as expected for a longitudinal mode. Finally, the radial breathing mode corresponds to 
\begin{eqnarray}\label{breath}
\omega_B(k) &=& \sqrt{ \frac{K+\mu}{\rho_M R^2} } + {\cal O}(k^2),\\ \nonumber
{\bf e}_B  (k) &=& \left(  \begin{array}{c}
0 \\  - i\nu k R \\ 1 \end{array}\right)  + {\cal O}(k^2).
\end{eqnarray}
This mode has an energy gap, $\hbar \omega_B \simeq 14$~meV for $R=1$~nm,
scaling as $\omega_B\propto R^{-1}$.   
The quoted results for the velocities $v_{T,L}$ and the frequency $\omega_B$,
first obtained in Ref.~\onlinecite{suzuura}, follow from Eq.~(\ref{lame}), and 
are in accordance with  {\em ab-initio} calculations.\cite{rubio}

For angular momentum $\ell=\pm 1$, we recover the correct 
dispersion relation of the important flexural ($J=F$) modes.\cite{mahan2}
They are degenerate and correspond to
\begin{eqnarray}\label{flex}
\omega_{F}^{}(k) &=& \frac{\hbar k^2 }{2m} + {\cal O}(k^4) 
, \quad m = \frac{\hbar}{\sqrt{2}\ v_L R} , \\ \nonumber
{\bf e}_{F,\ell=\pm}(k) &=& \frac{1}{\sqrt{2}} \left( 
\begin{array}{c}
 1 + \frac{(2\nu-1)k^2 R^2}{4} \\ 
\mp kR \left(1- \frac{( 9 + 6 \nu) k^2 R^2}{4} \right) \\
\mp i \left( 1 - \frac{(2\nu +1)k^2 R^2}{4} \right)  \end{array}
\right) + {\cal O}(k^4).
\end{eqnarray}
Note that for $|k|R \leq \sqrt{2}v_T/v_L$, and thus for all
wavelengths of interest here,  the flexural phonons are the 
lowest-lying modes available.  

Next we observe that for $\ell\ne 0$,  longitudinal modes
acquire a gap, $\omega_{L}(k=0,\ell) = v_T |\ell|/R$, and 
``breathing'' modes have an even larger gap
than Eq.~(\ref{breath}), $\omega_B(0,\ell)= \sqrt{1+\ell^2} \omega_B(0,0)$.
Since we focus on low-energy acoustic modes,
these gapped modes are irrelevant and will not be studied further.
Moreover, the diagonalization of the elastic matrix (\ref{elmat}) 
shows that for any $\ell$ \textit{flexural}\ modes
remain gapless. However, for $|\ell| > 1$, curvature effects
(see App.~\ref{appA}) will open gaps for these modes as well.\cite{suzuura} 
For $R\alt 1$~nm, such gaps are comparable in magnitude 
(or slightly smaller than)
the frequency of the breathing mode (\ref{breath}).\cite{suzuura} 
Since ph-ph interaction effects become more and more pronounced 
with decreasing radius $R$ (see below), the most interesting 
application range of our theory is 4~\AA$\alt R\alt 1$~nm, 
where most phonon modes have rather large gaps but a continuum
elasticity approach is still reliable.  Ignoring gapped modes is 
then a good approximation over a wide temperature regime 
and in our low-energy approach we need to retain only 
gapless modes, i.e.~the $T$ mode (\ref{twist}), the $L$ mode (\ref{long}), 
and the two degenerate flexural $F$ modes (\ref{flex}).  
The gaplessness of the $\ell=\pm 1$ flexural modes is robust 
against curvature effects and protected by rotational symmetry.
Note that the resulting theory is only valid 
on energy scales below those gaps;  
for $R\approx 0.5$~nm, this is justified up to temperatures of order of $50$~K. 

{}From now on, the sums over $(J,\ell)$ will then only run over 
$(T,0), (L,0)$, and $(F,\pm)$.  It is remarkable that for all these 
phonon modes, elastic continuum theory is able to yield accurate 
dispersion relations which are in good agreement with elaborate 
force-constant\cite{mahan2,jiang,white,zimmermann}
and \textit{ab-initio} calculations.\cite{rubio}
Since the breathing mode (\ref{breath})
may be of interest for future thermal expansion calculations,
we specify the corresponding three-phonon matrix elements 
in App.~\ref{appB}, but for the main part of the paper 
we will neglect this mode.  
 
\section{Phonon-phonon interaction processes}\label{sec3}

In this section, we evaluate the three- and four-phonon
scattering amplitudes following from Eqs.~(\ref{h3}) and (\ref{h4}),
respectively.  They are obtained by inserting the normal-mode expansion 
(\ref{displ}) for the displacement field into the definition
of the strain tensor, see Eqs.~(\ref{ulin}) and (\ref{unlin}).
We will always keep the lowest nontrivial order in $|k|R\ll 1$, 
but also specify the next order when cancellation effects 
are anticipated for the leading order.  It is then straightforward 
to obtain the full normal-mode representation of
the nonlinear strain tensor.  The result can be found in explicit form 
in Appendix \ref{appB}. 

\subsection{Three-phonon processes}

The normal-mode representation of the strain tensor
in App.~\ref{appB} allows us to write $H^{(3)}$ from Eq.~(\ref{h3}) 
in the form of a standard 
 three-phonon interaction Hamiltonian\cite{maradudin,cowley} (note again that
$\sum_k=\int_{-\infty}^\infty \frac{dk}{2\pi}$),
\begin{eqnarray}\label{3phonon}
H^{(3)}&=& \frac{1}{\sqrt{2\pi R}} \sum_{J_1 J_2 J_3} \sum_{k_1 k_2} 
A_{J_1 J_2 J_3}(k_1,k_2,k_3) 
\\ \nonumber
&\times&  u_{J_1}(k_1) u_{J_2}(k_2) u_{J_3}(k_3) , 
\end{eqnarray} 
where the $\ell$ summation is implicit when $J=F$, i.e.~$J$ stands
for both the phonon mode index and the angular momentum $\ell$.
Due to momentum conservation $k_3=-k_1-k_2$,
and  $u_{J}(k, \ell)$ has been defined in Eq.~(\ref{unewdef}).  
After some algebra, we obtain the following 
\textit{non-vanishing three-phonon amplitudes}
$A_{J_1 J_2 J_3}(k_1,k_2,k_3)$ to leading order in $|k_i|R\ll 1$,
\begin{eqnarray}\label{lll}
A_{LLL} &=& - \frac{i}{2} (1-\nu)k_1 k_2 k_3 [ 2K  \\ \nonumber &+& 
(1-\nu)^2 \xi_1 + (1+\nu^2)\xi_2 ], \\ 
\label{ltt} A_{LTT} &=& -\frac{i}{2} \Bigl( [ 2\mu +\xi_2 (1-\nu) ]k_1 k_2 k_3
\\ \nonumber 
& - & \nu \mu k_1^3\Bigr), \\ \label{lff}
A_{L,F\ell_2,F\ell_3}  &=& -i \delta_{\ell_2,-\ell_3} \mu (1+\nu) k_1 k_2 k_3,
\\ \label{tff}
A_{T,F\ell_2,F\ell_3} &=& -\frac{i \ell_2 \mu}{4} \delta_{\ell_2,-\ell_3} 
 k_1 k_2 k_3 (k_2-k_3)R,
\end{eqnarray}
with $\nu$ in Eq.~(\ref{poisson}).
The matrix elements related to the breathing mode can be found in
App.~\ref{appB}. Symmetry under phonon exchange is 
taken into account in the expressions (\ref{lll})--(\ref{tff}),
and the $(J_1,J_2,J_3)$ summation in Eq.~(\ref{3phonon}) runs only over 
$(LLL), (LTT), (LFF)$ and $(TFF)$, while all other matrix
elements vanish identically. In particular, there is no 
amplitude for $LLT$ processes\cite{mingo2} nor for the scattering of
three twist modes, $TTT$.  Moreover,
all amplitudes involving an odd number of flexural phonons vanish
by angular momentum conservation.  We also observe that the
anharmonic third-order couplings $\xi_1$ and $\xi_2$ do not
introduce new physics, but only renormalize parameter values
of coupling terms generated already by geometric 
nonlinearities. 
%(For instance, $A_{LLL}=-i C k_1 k_2 k_3$ with $C=(1-\nu)K$
%for $\xi_{1}=\xi_2=0$.  Finite anharmonic third-order couplings $\xi_{1,2}$
%then modify the value of $C$ but do not change the structure of $A_{LLL}$.)
In fact, the leading contributions to phonon
decay rates turn out to be completely independent of such 
anharmonic couplings, as we will show in Sec.~\ref{sec4},
see Eqs.~(\ref{gamml}), (\ref{gammt}) and (\ref{finalrate}) below.

\subsection{Four-phonon processes and flexural phonon interaction}
\label{sec3.4}

Next we turn to four-phonon interactions. 
A similar result as for three-phonon interactions, see Eq.~(\ref{3phonon}),
can be derived using the strain tensor given in App.~\ref{appB}. 
Since $FFF$ matrix elements vanish, quartic terms are crucial in the 
case of flexural modes, and we shall only discuss these
four-phonon matrix elements in what follows.  It is nevertheless 
straightforward (if tedious) to study also other four-phonon matrix 
elements based on the expressions given in App.~\ref{appB}.  

Since the flexural mode is the lowest-lying phonon 
branch, $FFFF$ processes provide the only possibility for its decay at $T=0$.
It turns out that the relevant coupling strength for such processes
is parametrized by 
\begin{equation}\label{gdef}
g = \frac{K+\frac{3}{16}(K+\mu)}{2\pi R}\left( \frac{m}{\rho_M}\right)^2,
\end{equation}
where $K=\mu+\lambda$ and $m$ is given in Eq.~(\ref{flex}).
Note that $g\propto 1/R^3$, and thus flexural phonon interactions 
 become stronger for thinner CNTs.  

After some algebra we find
\begin{eqnarray*}
H^{(4)} &=& \frac{1}{2\pi R} \sum_{k_1 k_2 k_3} \sum_{\{\ell\}}
A_{F\ell_1,F\ell_2,F\ell_3,F\ell_4} (k_1,k_2,k_3,k_4) \\
&\times& 
\ u_{F}(k_1, \ell_1) u_{F}(k_2,\ell_2)  u_{F}(k_3,\ell_3) u_{F}(k_4,\ell_4) ,
\end{eqnarray*}
where $k_4=-(k_1+k_2+k_3)$, angular momentum conservation implies the condition
$\ell_1+\ell_2+\ell_3+\ell_4=0$, and 
\begin{eqnarray*}
&& A_{F\ell_1,F\ell_2,F\ell_3,F\ell_4}  = 
 \frac{k_1k_2k_3 k_4}{8} \\
&& \times
\left( K - \frac{K}{6} \sum_{i<j} \ell_i\ell_j + \frac{K+\mu}{4}\prod_{i=1}^4
\ell_i \right).
\end{eqnarray*}
Now for all $\{\ell_i\}$ combinations with $\sum_{i=1}^4\ell_i=0$ and
$\ell_i=\pm 1$, one finds $\sum_{i<j}\ell_i\ell_j = -2$ and $\prod_i \ell_i=1$. 
This allows us to carry out the $\ell$ summation, and gives
\begin{eqnarray}\label{4phonon}
H^{(4)} &=& g (\rho_M/m)^2 \sum_{k_1 k_2 k_3}
k_1 k_2 k_3 k_4 \\ \nonumber
&\times&   u_{F}(k_1,+) u_{F}(k_2,+) u_{F}(k_3,-) u_{F}(k_4,-).
\end{eqnarray}
Note that Eq.~(\ref{4phonon}) is determined by geometric nonlinearities
alone, i.e.~by the contribution of the nonlinear part of the strain tensor in ${\cal U}_2[u]$.
The anharmonic third- and fourth-order couplings
($\xi_{1,2}$ and $\kappa_{1,2,3}$, respectively) also give rise
to $A_{FFFF}$ contributions, which however contain higher powers
in $|k_i| R \ll 1$.  Such anharmonic four-phonon processes are therefore
parametrically smaller than the geometric nonlinearity (\ref{4phonon}),
and can be neglected in a low-energy approach.

In coordinate space,
Eq.~(\ref{4phonon}) corresponds to a \textit{local}\ four-phonon interaction.
To see this, we represent the momentum conservation constraint in 
Eq.~(\ref{4phonon}) as
\[
\delta(k_1+k_2+k_3+k_4) = \int \frac{dy}{2\pi} e^{-i(k_1+k_2+k_3+k_4)y} ,
\]
and then arrive at 
\begin{eqnarray}\label{4ph2}
H &=& \int dy \Biggl[ \frac{p^\dagger(y) p(y)}{\tilde \rho_M} + 
\frac{\hbar^2\tilde \rho_M}{4m^2} \frac{\partial^2 u^\dagger(y)}{\partial y^2}
 \frac{\partial^2 u(y)}{\partial y^2} \\ \nonumber &+&
 g (\tilde\rho_M/m)^2 \left(\frac{\partial u^\dagger}{\partial
y} \frac{\partial u}{\partial y}\right)^2 \Biggr],
\end{eqnarray}
where $\tilde \rho_M=2\pi R \rho_M$ is the effective linear mass density,
and  the (non-Hermitian) coordinate-space flexural displacement operator is
defined as
\begin{eqnarray}\label{uudef}
u(y) &=& \frac{1}{\sqrt{2\pi R}} \sum_k e^{iky} u_F(k,+) \\ \nonumber
& =& \sum_k \sqrt{\frac{\hbar}{2\tilde \rho_M \omega_F(k)}}
 e^{iky} \left( a^{}_{F}(k,+)+a^\dagger_F(-k,-) \right),
\end{eqnarray}
with the canonically conjugate momentum field operator
\begin{eqnarray*}
p(y) &=&\sqrt{2\pi R}\sum_k e^{iky} p_F(k,-)
= -i \sum_k \sqrt{\frac{\hbar \tilde \rho_M\omega_F(k)}{2}} \\&\times&
e^{iky} \left( a^{}_F(k,-)-a^\dagger_F(-k,+) \right).
\end{eqnarray*}
Since $g>0$, the interaction among flexural phonons is \textit{repulsive}. 
Therefore phonon localization and two-phonon bound states\cite{kimball} 
are not expected to occur.

Remarkably, as we show in detail below, it turns out that the
finite-temperature decay rate $\Gamma_F(k)$ 
for a flexural phonon  \textit{diverges} when the interaction (\ref{4phonon}) 
is treated perturbatively.  A related breakdown of perturbation 
theory for phonon decay rates has also been reported by Perrin\cite{perrin}
in a study of optical phonons in molecular crystals.
In that case, the singularity could be traced to the flatness of the 
dispersion relation. 
A similar situation occurs for the magnon decay problem in 1D spin chains,
where the analogous perturbation theory also predicts a 
finite and momentum-independent $T=0$ decay rate above a certain threshold, 
while the correct (nonperturbative) result vanishes at the thresholds.\cite{essler}
In our case, the divergence arises due to the conspiracy of the almost flat
dispersion relation, $\omega_F(k)=\hbar k^2/2m$, with
the low dimensionality (1D). This implies a \textit{macroscopic} 
phonon generation in the noninteracting case for finite $T$.
For a system of length ${\cal L}$, the total number
of $\ell=\pm$ flexural phonons follows with the 
Bose-Einstein distribution function ($\beta=1/k_B T$),
\begin{equation}\label{boseeinstein}
n(\omega)=\frac{1}{e^{\beta \hbar \omega}-1},
\end{equation}
as $N=2{\cal L}\sum_k n(\omega_F(k))$.  As a result, the  1D phonon density
 $\rho= N/{\cal L}\approx 2 m k_B T {\cal L}/\pi \hbar^2$ diverges in the 
thermodynamic limit ${\cal L}\to \infty$ at any finite temperature $T$.
This situation therefore calls from the outset for a nonperturbative 
treatment of the interaction (\ref{4phonon}).  In view of the divergent
noninteracting phonon density $\rho$, we expect that mean-field theory 
is able to properly handle the regularizing effect of the interaction 
despite the low dimensionality, at least in a semi-quantitative fashion.  
We thus employ mean-field theory to compute the 1D flexural phonon 
density $\rho(T)$, and then use this result in Sec.~\ref{sec4}
for the decay rate calculations.

Taking $\bar n_{k\ell}=\langle a_F^\dagger(k,\ell) a^{}_F(k,\ell)\rangle$ 
as the only non-vanishing mean-field parameters in Eq.~(\ref{4phonon}),
the mean-field Hamiltonian for the flexural modes
is given (up to irrelevant constants) by
\begin{equation}\label{mf}
H_{\rm MF}= \sum_{k,\ell=\pm}\ [\hbar \omega_F(k) + 4g\rho] \ 
a^\dagger_F(k,\ell) a^{}_F(k,\ell) ,
\end{equation}
where $\rho=\sum_{k\ell} \bar n_{k\ell}$.
In order to derive Eq.~(\ref{mf}), we disregard all terms involving
an unequal number of creation ($a$) and annihilation ($a^\dagger$)
operators, see also Ref.~\onlinecite{kimball}.  Moreover, for
all nonvanishing contributions to the 
mean-field approximation of Eq.~(\ref{4phonon})
one finds ${\rm sgn}(k_1 k_2 k_3 k_4)=1$. 
The resulting self-consistency equation is then
$\rho = 2 \sum_k n( \omega_F(k)+4g\rho )$.
The momentum integral can be carried out and yields
\begin{equation}\label{meandens}
\rho = \frac{k_B T^*}{4g} (T/T^*)  Y( T/T^* ).
\end{equation}
The temperature dependence of $\rho$ shows universal scaling
with $x=T/T^*$, where we introduce the temperature scale
\begin{equation}\label{tstar}
T^* =\frac{32 m g^2}{k_B\hbar^2} = \frac{16\sqrt{2} g^2}{k_B \hbar v_L R }.
\end{equation}
The dimensionless scaling function $Y(x)$ is determined by the 
self-consistency condition 
\begin{equation}\label{yself}
\sqrt{\pi x} \ Y =  {\rm Li}_{1/2}\left(e^{-Y}\right),
\end{equation}
with the polylogarithm\cite{gr}
${\rm Li}_s(z) = \sum_{j=1}^\infty j^{-s} z^j$.
Equation (\ref{yself}) can be analytically solved in the limits
$x\ll 1$ and $x\gg 1$, and allows for numerical evaluation in between
those limits.  In particular, we find $Y(x\ll 1)\simeq
-\frac12 \ln(\pi x)$ and $Y(x\gg 1)\simeq x^{-1/3}$,
where we exploit the relation\cite{gr} $\lim_{Y\to 0} {\rm Li}_{1/2}(e^{-Y})
=\sqrt{\pi/Y}$.  In accordance with our above discussion, 
we therefore find that 
in the noninteracting  ($T^*=0$) case, $\rho$ diverges for any finite $T$.  
However, once interactions are present, a finite flexural phonon 
density $\rho$ emerges, which for $T\gg T^*$ can be written as
\begin{equation}\label{rhofinal}
\rho (T\gg T^*) = \frac{k_B T^*}{4g}  (T/T^*)^{2/3}.
\end{equation}
Using the parameters in Eq.~(\ref{lame}),
the scale $T^*$ in Eq.~(\ref{tstar}) is estimated as 
\begin{equation}\label{tstarestimate}
T^{*} \simeq \frac{3.7\times 10^{-9}~{\rm K}}{(R [{\rm nm}])^7}.
\end{equation}
Even for the thinnest possible CNTs (where $R\approx 0.3$~nm),  
this puts $T^*$ deep into the sub-milli-Kelvin regime.
Assuming $T\gg T^*$ from now on, we take $\rho$ as given 
in Eq.~(\ref{rhofinal}). Within mean-field theory,
 see Eq.~(\ref{mf}), nonperturbative effects of the 
interaction (\ref{4phonon}) thus lead to the appearance of a dynamical
gap $\omega_\rho=4g\rho$ for flexural phonons.  We effectively arrive at a 
\textit{modified flexural dispersion relation},
\begin{equation}\label{flexnew}
\omega_F(k) = \omega_\rho + \frac{\hbar k^2}{2m},
\end{equation}
characterized by the temperature-dependent gap 
\begin{equation}\label{wr}
\hbar\omega_\rho(T) =  (T/T^*)^{2/3} k_B T^*.
\end{equation} 
{}From now on, we take Eq.~(\ref{flexnew}) for the dispersion 
relation of flexural phonons.  Since $T\gg T^*$, we also observe
that the gap is always thermally smeared, $k_B T\gg \hbar\omega_\rho$.
Nevertheless, it is crucial when discussing the decay rate
for a flexural phonon.

\section{Decay rate and quality factor of acoustic phonon modes}
\label{sec4}

In this section, we study the decay rate of a phonon excitation with
longitudinal momentum $p=\hbar k>0$ and mode index $J=L,T$ or $F$.  
We compute the finite-temperature decay rate $\Gamma_{J}(k,\ell)$
from lowest-order perturbation theory in the relevant nonlinearity. 
At $T=0$, this corresponds to the standard Fermi's golden rule result.  

\subsection{Self-energy calculation}

To access the finite-$T$ case, we will first write down the respective
imaginary-time self-energy $\tilde\Sigma_{J}(\tau,k,\ell)$, 
where $0\leq \tau<\hbar\beta$ denotes imaginary
time.  The Matsubara Green's function is defined via
\begin{eqnarray*}
 && \langle u_{J}(\Omega_n,k,\ell) u_{J}(-\Omega_n',-k',-\ell)\rangle = 
\\ &&  - 2\pi \hbar \beta \delta_{\Omega_n,\Omega'_n} 
\delta(k-k') \tilde G_{J}(\Omega_n,k,\ell),
\end{eqnarray*}
where the $\Omega_n=2\pi n /\hbar\beta$ (integer $n$)  are bosonic Matsubara 
frequencies and $u_{J}(\tau,k,\ell)$ is defined in Eq.~(\ref{unewdef}). 
Employing Eqs.~(\ref{nonint}) and (\ref{unewdef}), the noninteracting 
Green's function is 
\begin{equation}\label{gf}
\tilde G^{(0)}_{J}(\Omega_n,k,\ell)  = \frac{-\hbar/\rho_M}{\Omega_n^2+
\omega_{J}^2(k,\ell)}= \frac{\hbar}{\rho_M}
 {\cal G}(\Omega_n,\omega_{J}(k,\ell)).
\end{equation}
The function ${\cal G}(\Omega_n,\omega_1)$ has the time-representation 
\begin{equation} \label{aux1}
{\cal G}(\tau,\omega_1) = - \frac{1}{\hbar \beta} \sum_{\Omega_n}
 \frac{e^{-i\Omega_n \tau}}{\Omega_n^2 + \omega_1^2} 
= -\frac{1}{2\omega_1} \sum_{\xi=\pm} \xi n(\xi \omega_1) e^{\xi\omega_1 \tau}
\end{equation}
with the Bose function (\ref{boseeinstein}).
The full retarded Green's function $G_{J}(\omega,k,\ell)$
follows after analytic continuation, $i\Omega_n\to \omega+i0^+$, 
with the Dyson equation,
\begin{equation}
G^{-1}_{J}(\omega,k,\ell) = (G^{(0)})^{-1}_{J}(\omega,k,\ell)-
 \Sigma_{J}(\omega,k,\ell), 
\end{equation}
leading to the on-shell $[ \omega=\omega_{J}(k,\ell) ]$ rate 
\begin{equation}\label{rate}
\Gamma_{J}(k,\ell)= \frac{\hbar}{\rho_M \omega} {\rm Im}\Sigma_{J}(\omega,k,\ell).
\end{equation}
The relevant quantity needed to estimate the decay rate is therefore the
self-energy $\Sigma_{J}(\omega,k,\ell)$, whose imaginary-time version
is $\tilde\Sigma_{J}(\Omega_n,k,\ell)$.

In the self-energy calculation for the various modes shown below, 
we will encounter integrals of the type (integer $r\ge 1$)
\begin{equation}\label{ir}
I_r (\Omega_n;\omega_1,\ldots,\omega_r) = \int_0^\beta d\tau 
e^{i\Omega_n\tau} \prod_{j=1}^r {\cal G}(\tau,\omega_j) .
\end{equation}
Employing Eqs.~(\ref{aux1}) and (\ref{boseeinstein}), we find for $r=2$ 
(see also Refs.~\onlinecite{maradudin,cowley})
\begin{equation}\label{i2}
I_2(\Omega_n,\omega_1,\omega_2)= \sum_{\xi_1,\xi_2=\pm} 
\frac{\xi_1\xi_2}{4\omega_1\omega_2} \frac{1+n(\xi_1\omega_1)+n(\xi_2\omega_2)}
{i\Omega_n+\xi_1\omega_1+\xi_2\omega_2}.
\end{equation}
Similarly, for $r=3$, we obtain\cite{perrin}
\begin{widetext}
\begin{equation}\label{i3}
I_3 (\Omega_n;\omega_1,\omega_2,\omega_3) =
-\frac{1}{8\omega_1\omega_2\omega_3} \sum_{\xi_1,\xi_2,\xi_3=\pm}
\frac{\xi_1\xi_2\xi_3}{i\Omega_n+\xi_1\omega_1 +\xi_2\omega_2+\xi_3\omega_3}
 \frac{n(\xi_1\omega_1) n(\xi_2\omega_2) n(\xi_3\omega_3)}{
n(\xi_1\omega_1+\xi_2\omega_2+\xi_3\omega_3)} .
\end{equation}
\end{widetext}
Let us then proceed with the discussion of the different phonon modes,
starting with $J=L$. 

\subsection{Longitudinal stretch mode}

The dominant contributions to the decay rate for a longitudinal phonon
come from the relevant non-vanishing three-phonon matrix
elements, namely $L\to L+L$ in Eq.~(\ref{lll}), 
$L\to T+T$ in Eq.~(\ref{ltt}), and $L\to F+F$ in Eq.~(\ref{lff}).
The amplitude for the $L\to L+T$ process vanishes,
and such decay channel could only be possible via higher-order
processes involving the virtual excitation of flexural
modes.  While one could compute the corresponding 
contribution, we expect that it is negligible against the rate found below. 
We also anticipate that the contribution of the process $L\to T+T$
is subleading with respect to that of $L\to F+F$, as dimensional arguments
at $T=0$ suggest and the explicit finite-$T$ calculation shows.
Therefore, we also neglect this decay channel.
Finally, while the process $L\to L+L$ is in principle kinematically 
allowed for a strictly linear dispersion relation, energy conservation
cannot be satisfied as soon as one takes into account the  ${\cal O}(k^2)$
corrections to $\omega_L(k)$. Thus, this decay channel can also be safely omitted. 

The only remaining possibility is then the process $L\to F+F$, where the 
two flexural phonons carry opposite angular momentum.
Energy conservation then poses no problem as long as $\omega_\rho\ll v_L k$,
see Eq.~(\ref{wr}).  For clarity, we now focus on this case, 
where the channel $L\to F+F$ provides the dominant decay 
mechanism for a $L$ phonon.
The lowest order in perturbation theory generating a finite decay rate 
comes from the ``bubble'' diagram (i.e.~the second order), 
\[
\tilde\Sigma_L(\tau,k)= \frac{4\hbar^2}{2\pi R \rho_M^2}\int 
\frac{dq}{2\pi}|A_{LFF}(k,q_1,q_2)|^2
{\cal G}(\tau,\omega_1) {\cal G}(\tau, \omega_2),
\]
where $q_{1,2}=\mp q+k/2$ and $\omega_{1,2}\simeq \hbar q_{1,2}^2/2m$,
and the amplitude $A_{LFF}$ in Eq.~(\ref{lff}) is evaluated, say,
for $\ell_2=-\ell_3=1$.  The two Green's functions correspond to flexural
phonons.  Using $\mu(1+\nu) = \rho_M v_L^2/2$, we then find 
\[
\tilde \Sigma_L(\Omega_n, k)= \frac{(\hbar v_L k)^2}{2\pi^2 R^3}
\int dq \ \omega_1\omega_2 I_2(\Omega_n;\omega_1,\omega_2),
\]
where $I_2$ is given in Eq.~(\ref{i2}).  For $k R\ll 1$, after 
analytic continuation, we obtain the rate from Eq.~(\ref{rate}). 
Identifying $\omega=v_L k$ yields with Eq.~(\ref{boseeinstein})
the result
\begin{eqnarray*}
&& \Gamma_L(k) = \frac{\hbar \omega}{4 \rho_M R^3} 
\int_{-\infty}^\infty \frac{ dq}{2\pi} \{ [n(\omega_1)+n(\omega_2)+1] \\
&&\times \delta(\omega-\omega_1-\omega_2) + 2 [n(\omega_2)-n(\omega_1)] 
\delta(\omega-\omega_1+\omega_2)\}.
\end{eqnarray*}
We now need to resolve the $\delta$-functions representing 
energy conservation.  The first term yields  
$q=\pm \sqrt{\frac{k}{R\sqrt{2}}-\frac{k^2}{4}}$,
while the second leads to $q=-\frac{1}{R\sqrt{2}}$.
We then collect terms, keeping only the leading order in $kR\ll 1$ and 
$k_B T \ll \hbar v_L/R$ -- otherwise the thermal scale $k_B T$ would 
exceed the smallest gap of the discarded phonon modes, and we 
would need to account for the effects of the mean-field gap 
 $\omega_\rho$. We find
\begin{eqnarray}\label{gamml}
\Gamma_L(k) &=& \frac{\hbar}{4\pi \rho_M R^4} 
\Biggl( \frac{\sqrt{kR}}{2^{5/4}} \coth(\hbar \beta v_L k/4)
\\ \nonumber && +
\sqrt{2} e^{-\frac{\beta \hbar v_L}{2\sqrt{2} R}} \sinh( \hbar\beta v_L k/2)
\Biggr).
\end{eqnarray}
Since this rate does not depend on the anharmonic couplings ($\xi_{1,2}$ 
and $\kappa_{1,2,3}$), we find the remarkable result
that the dominant decay rate for the longitudinal phonon  
is solely determined by the Lam{\'e} coefficients 
(or, equivalently, by the sound velocities).
For $T=0$ and $k\to 0$, the rate becomes {\em universal}, i.e. completely 
independent of material parameters. This is due to the fact that in 
our elastic model the curvature of the flexural branch $\sim 1/m$ 
is proportional  to the longitudinal sound velocity $v_L$.
Moreover, the $T=0$ rate is $\propto \sqrt{k}$, i.e. 
the nonlinear damping effects become important at 
long wavelengths. 

For an estimate of the \textit{quality factor} for 
longitudinal modes, we now put $k=\pi/{\cal L}$ in 
$Q_L=\omega_L(k)/\Gamma_L(k)$, 
with CNT length ${\cal L}$.  This yields
\begin{widetext}
\begin{equation} \label{qualL}
Q_L(T,{\cal L}) = \frac{Q_L(0,{\cal L})}
{ \coth\left(\frac{\pi \hbar v_L }{4{\cal L} k_B T}\right)
+ 2^{7/4} \sqrt{\frac{{\cal L}}{\pi R}} e^{-\frac{\hbar v_L}{2\sqrt{2} R k_B T}}
\sinh\left( \frac{ \pi \hbar v_L }{2 {\cal L} k_B T}\right)},
\end{equation}
\end{widetext}
with the $T=0$ result
\begin{equation}\label{zerotqualL}
Q_L(0,{\cal L}) = (3.80 \times 10^6)  \times (R [{\rm nm}])^3 \sqrt{R/{\cal L}},
\end{equation}
where we used Eq.~(\ref{lame}).  For typical parameters, say, 
$R=0.5$~nm and ${\cal L}=500$~nm, this gives $Q_L(T=0)\approx 1.5\times 10^4$. 

This number for the quality factor implies a surprisingly 
strong damping effect due to phonon-phonon interactions, and is 
similar to what is observed experimentally.\cite{bachtold}
Note that this value yet has to be 
understood as upper bound since there might be other decay mechanisms. 
Since $\omega_\rho(T=0)=0$, inclusion of the mean-field gap $\omega_\rho$ in 
the above derivation does not affect the estimate (\ref{zerotqualL}).
Furthermore, with $Q_L(0,{\cal L})\propto {\cal L}^{-1/2}$, we observe that
for sufficiently long CNTs and low $T$, the anharmonic decay is 
always important, in agreement with the conclusions of 
Ref.~\onlinecite{mingo1}, see our discussion
in Sec.~\ref{sec1}.  The temperature dependence of $Q_L$ is controlled 
by the ratio of the longitudinal confinement energy scale, 
$\hbar v_L/{\cal L}$, and the thermal energy, $k_B T$.
For $k_B T\gg \hbar v_L/{\cal L}$, we find 
\begin{equation}\label{hight}
Q_L(T,{\cal L})\approx \frac{\pi \hbar v_L}{4k_B T {\cal L}} Q_L(0,{\cal L})
\propto \frac{R^{7/2}}{{\cal L}^{3/2} T}.
\end{equation}
We observe that the damping effects get stronger for thinner CNTs.

\subsection{Twist mode}
 
Next we turn to the twist ($J=T$) mode, where the vanishing of the
$T\to T+T$ amplitude and the absence of the (kinematically forbidden)
$T\to T+L$ channel imply that the $T\to F+F$ decay will dominate.
The calculation proceeds in the same way as for the
$L$ phonon, and in the low-energy limit, $k_B T
\ll \hbar v_T/R$ and $\omega_\rho\ll v_T k$, we find
\begin{widetext}
\begin{eqnarray}\label{gammt}
\Gamma_T(k) &=& \frac{\hbar}{2\rho_M} (\frac{v_T}{v_L})^{7/2} 
\frac{2^{1/4}(kR)^{3/2}}{8\pi R^4} \Biggl( \coth\left(\frac{\hbar 
v_T k }{4 k_B T}\right) \\ && \nonumber
+ 2^{5/4} [(v_L/v_T) kR]^{-3/2} e^{-\frac{\hbar v_T^2}{2\sqrt{2}v_L R k_B T}}
\sinh\left( \frac{\hbar v_T k}{2k_B T}\right) \Biggr).
\end{eqnarray}
We note that  the $T=0$ rate is $\propto k^{3/2}$.
\end{widetext}
Putting $k=\pi/{\cal L}$, the quality factor is then 
\begin{equation}
Q_T(T,{\cal L})=\frac{\pi v_T/{\cal L}}{\Gamma_T(T,{\cal L})}.
\end{equation}
This gives for $T=0$ the estimate
\begin{equation}
Q_T(T=0,{\cal L}) \simeq (5.74\times 10^6) \times (R [{\rm nm}])^3 \sqrt{{\cal L}/R},
\end{equation}
showing that damping of the twist mode, which is energetically
below the $L$ mode, $v_L/v_T\simeq 0.62$, is much weaker, in accordance
with Ref.~\onlinecite{lax}.  In the high-temperature limit, we find
$Q_T\propto R^{5/2}/(T {\cal L}^{1/2})$. 

\subsection{Flexural mode}

\begin{figure}
\scalebox{0.5}{\includegraphics{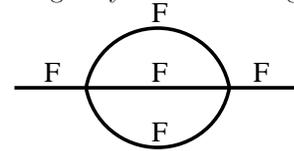}}
\caption{\label{fig1} Self-energy diagram contributing 
to the decay rate $\Gamma_F(k)$ in leading order 
(``fishbone diagram''). Solid lines denote a flexural 
phonon propagator, taken from mean-field theory.}
\end{figure}

Next we discuss the decay rate $\Gamma_F(k)$
for the flexural mode.  We put $\ell=+1$ (the rate for $\ell=-1$ is
identical), and describe the perturbative calculation of $\Gamma_F(k)$.
The leading term involves the decay
of the phonon into three flexural phonons, $F\to F+F+F$, 
see Eq.~(\ref{4phonon}), corresponding to 
the ``fishbone'' diagram for the self-energy in Figure 
\ref{fig1}.  Note that we have 
already taken into account the self-consistent
tadpole (first-order) diagram by using mean field theory, leading
to the dispersion relation (\ref{flexnew}).
The imaginary-time self-energy corresponding to the
fishbone diagram involves three Green's functions, with momenta $q_i$
and frequencies $\omega_i=\omega_\rho+ \hbar q_i^2/(2m)$, where $i=1,2,3$. 
Employing Eq.~(\ref{i3}) and the interaction strength $g$
in Eq.~(\ref{gdef}), we then obtain
\begin{widetext}
\begin{eqnarray}\label{self1}
\tilde \Sigma_F(\Omega_n,k) &=& \frac{\rho_M \hbar g^2 k^2 }{4 \pi^2 m^4}
\int dq_1 dq_2 dq_3 \frac{q_1^2 q_2^2 q_3^2}{\omega_1\omega_2\omega_3} 
\delta(q_1+q_2+q_3-k)\\ \nonumber
&\times& \sum_{\xi_1,\xi_2,\xi_3=\pm} 
\frac{\xi_1\xi_2\xi_3}{i\Omega_n+\xi_1\omega_1
+\xi_2\omega_2+\xi_3\omega_3}
 \frac{n(\xi_1\omega_1) n(\xi_2\omega_2) n(\xi_3\omega_3)}{
n(\xi_1\omega_1+\xi_2\omega_2+\xi_3\omega_3)}.
\end{eqnarray}
The $\delta$-function represents momentum conservation.
After analytic continuation,
using the relation $2n(\pm\omega) = \pm \coth(\hbar\beta\omega/2)-1$,
this yields the rate
\begin{eqnarray}\label{flexuralrate}
\Gamma_F(k) &=& \frac{\hbar^2 g^2 k^2}{2^5 \pi \omega m^4 [1+n(\omega)]} 
\sum_{\xi_1,\xi_2,\xi_3} \int dq_1 dq_2 dq_3 \, \delta(q_1+q_2+q_3-k) \\
&\times& \delta(\omega+\xi_1\omega_1+\xi_2\omega_2+\xi_3\omega_3)
\prod_{i=1}^3 \left( \frac{q_i^2}{\omega_i} [\coth(\hbar\beta\omega_i/2)-\xi_i]
\right). \nonumber
\end{eqnarray}
\end{widetext}
We now express the $\delta$-functions as
\begin{eqnarray*}
\delta\left(k-\sum_i q_i\right) &=&\int 
\frac{dy}{2\pi} e^{-i\left(k-\sum_i q_i\right)y}, \\
\delta\left(\omega+\sum_i\xi_i\omega_i\right)
&=& \int \frac{dt}{2\pi} e^{i\left(\omega+\sum_i\xi_i\omega_i\right)t},
\end{eqnarray*}
which decouples the $q_i$ integrals and allows to perform the $\xi_i$
summations. Setting $\omega=\omega_F(k)$, the on-shell rate reads
\begin{equation}\label{frate}
\Gamma_F(k)= \frac{8 g^2 (1-\omega_\rho/\omega)}{\hbar^2[1+n(\omega)]} 
\int dt dy \,e^{-i[ky -\omega t]} G^3(t,y) ,
\end{equation}
with the correlation function 
\begin{eqnarray} \label{flexrate}
 G(t,y) &=& \frac{\tilde  \rho_M}{m} \int_{-\infty}^\infty 
\frac{dq}{2\pi} \frac{\hbar q^2 \cos(qy)}{2\tilde \rho_M \omega_F(q)} \\
\nonumber &\times&\left( \cos[\omega_F(q) t] \coth[\hbar \beta\omega_F(q)/2] -
i \sin[\omega_F(q) t]\right )\\ &=&  \nonumber
\frac{\tilde \rho_M}{m} \langle [\partial_y u^\dagger](t,y) 
[\partial_y u](0,0) 
\rangle_0 .
\end{eqnarray}
Here $u(t,y)$ is the Heisenberg representation of the flexural 
displacement operator $u(y)$, see Eq.~(\ref{uudef}), and the 
noninteracting average $\langle\cdots\rangle_0$ is taken 
with respect to $H_{\rm MF}$,
see Eq.~(\ref{mf}).  The product of three Green's functions
in Eq.~(\ref{frate}) reflects the structure of the
fishbone diagram in Fig.~\ref{fig1}. 
Using the dispersion relation (\ref{flexnew}) and $\omega_\rho$ 
in Eq.~(\ref{wr}), we observe that the $q$-integral for $G(t,y)$ 
is regular.

The rate $\Gamma_F$ for the decay of the mode 
with wavevector $k=\pi/{\cal L}$
then depends only on the two dimensionless quantities
\begin{equation}\label{scale}
X_{\cal L}=\frac{\cal L}{{\cal L}^*},\quad X_T = \frac{T}{T^*}, 
\end{equation}
where we define the lengthscale 
\begin{equation}\label{ldef}
{\cal L}^*= \frac{\pi\hbar}{\sqrt{2m k_B T^*}} = \frac{\pi \hbar^2}{8 mg}.
\end{equation} 
Using Eq.~(\ref{tstarestimate}), we obtain the estimate
\[
{\cal L}^* [\mu{\rm m}] \simeq 534 \ (R [{\rm nm}])^4, 
\]
which gives ${\cal L}^*=33.4~\mu$m for $R=0.5$~nm.  
By rescaling all lengths in units of ${\cal L}^*$ and
all frequencies (or inverse times) in units of $k_B T^*/\hbar$,
Eqs.~(\ref{frate}) and (\ref{flexrate}) imply a \textit{universal} result,
where the dependence on material parameters only enters via $T^*$
and ${\cal L}^*$.
Using Eqs.~(\ref{flexnew}) and (\ref{wr}), after some algebra, we  obtain 
\begin{widetext}
\begin{eqnarray} \label{finalrate}
 && \frac{\hbar \Gamma_F(X_{\cal L},X_T)}{k_B T^*} = 
 \frac{1}{2} \frac{1-\exp\left(-\frac{1}{X_T^{1/3}} -
 \frac{1}{X_T X_{\cal L}^2}\right)}
 {1+ X_T^{2/3} X_{\cal L}^2} 
\int dy dt \exp\left(- i \left [\frac{y}{X_{\cal L} }- \left(
X_T^{2/3}X_{\cal L}^2 +1\right) \frac{t}{X_{\cal L}^2}\right]\right)  
\\ \nonumber &&\times \left( \int_{X_T^{2/3}}^\infty 
\frac{dw}{2\pi w} \sqrt{w-X_T^{2/3}}  \ \cos\left(\sqrt{w-X_T^{2/3}} \ 
y\right) \ \left [ \cos(w t)\coth(w/2X_T)-i\sin(w t) \right] \right)^3.
\end{eqnarray}
\end{widetext}

For $T=0$, Eq.~(\ref{finalrate}) can be solved in closed form and gives 
\begin{equation}\label{flexrate0}
\Gamma_F(X_{\cal L}, X_T=0) = \frac{k_B T^*}{4\sqrt{3}\hbar}.
\end{equation}
(In fact, this result easily follows also from Eq.~(\ref{flexuralrate}).)
Remarkably, this rate does not depend on the length ($X_{\cal L}$) of the
CNT, i.e., it is independent of phonon momentum $k$.
Despite the smallness of $T^*$, see the estimate in
 Eq.~(\ref{tstarestimate}), this predicts a surprisingly large damping effect
due to phonon interactions.  Estimating the zero-temperature
quality factor as above, we find
\begin{equation}
Q_F(T=0,{\cal L}) \simeq (1.98\times 10^{12}) 
\times (R \ [{\rm nm}])^6 (R/{\cal L})^2.
\end{equation}
Taking ${\cal L}=500$~nm and $R=0.5$~nm, 
this gives $Q_F\approx 3\times 10^4$.

\begin{figure}
\scalebox{0.35}{\includegraphics{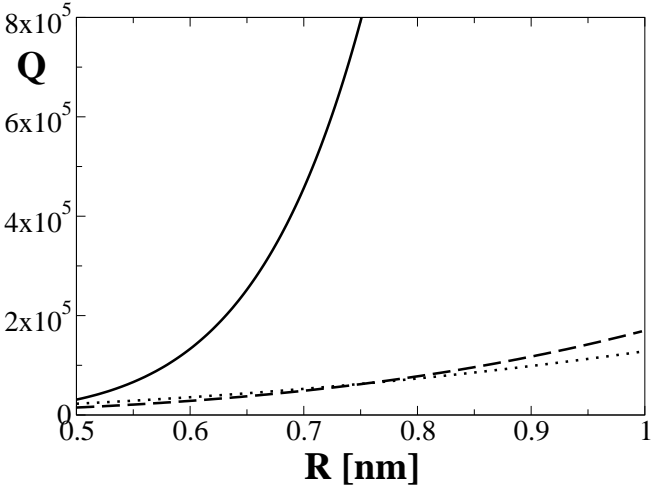}}
\caption{\label{fig2} 
Quality factor for the three low-energy modes 
at $T=0$ and ${\cal L}=500$~nm as a function of tube radius $R$. 
The solid line depicts $Q_F$, the dashed line 
$Q_L$ and the dotted line $10^{-3}Q_T$.
}
\end{figure}

The $R$-dependence of the zero-temperature quality factors 
for the various modes is summarized in Fig.~\ref{fig2}. 
In the range of radii considered, it is much stronger for  
the flexural mode than for the longitudinal and twist modes, 
which have an approximately similar dependence. 
Note however that $Q_T$ is in fact  three orders of magnitude 
larger than $Q_L$.

However, as discussed in Sec.~\ref{sec3.4}, due to the smallness of $T^*$,
the zero-temperature limit is actually not accessible experimentally,
and one is always in the regime $T\gg T^*$. It is then interesting to 
evaluate the rate in the limit $X_T\gg 1$ but with $X_T^{1/3} X_{\cal L}\ll 1$,
which corresponds to a regime in which the kinetic energy is much larger than the
gap. For sufficiently short CNTs (or if one evaluates the rate at a larger wavevector
than $\pi/{\cal L}$) the two inequalities can be satisfied simultaneously.
In that case, we find from Eq.~(\ref{finalrate})
 that $\hbar \Gamma_F/k_B T^*\propto X_{\cal L}^3 X_T^2$.
The resulting $T^2$ power-law behavior for the temperature 
dependence of the flexural phonon decay rate is a prediction
that should be observable with state-of-the art experiments.   

\section{Electron-phonon coupling}\label{sec5}

In this section, we consider the effects of e-ph couplings in
\textit{metallic} single-wall CNTs.  In contrast to the semiconducting
case studied before, the coupling of phonons  to electrons
may contribute another decay channel beyond ph-ph interactions.  
Within lowest order perturbation theory, the two mechanisms are additive for 
the decay rates, and hence also for the inverse quality factors. 
Here, we only take into account the \textit{deformation potential} 
as a source for e-ph coupling, as this coupling has a very
significant strength.\cite{suzuura,ademarti} 
The phonons, described by the strain tensor $u_{ij}$, create
the deformation potential\cite{suzuura}
\begin{equation}\label{defpot}
V_{e-ph}({\bf r})= V_0 {\rm Tr}\left
[u^{\rm lin}({\bf r})+ u^{\rm nlin}({\bf r})\right],
\end{equation}
where $V_0\approx 20$~eV (see Ref.~\onlinecite{suzuura}). 
We will consider spinless electrons at a single Fermi point and
eventually multiply by a factor 4 the final result for the rate, 
to take into account the electron spin and $KK'$ degeneracies.
Moreover, we only take into acccount electrons in the lowest transverse 
subband, with zero angular momentum. This is justified since
the higher angular momentum states are separated by a large energy gap. 
Then, from the normal-mode representation in Appendix \ref{appB}, we observe
that only the $L$ mode couples via the first term  ($u^{\rm lin}$) in Eq.~(\ref{defpot})
(the coupling
to the $F$ mode requires at least one electron in a higher angular momentum state),
while the $F$ and $T$ phonons couple only 
via the second ($u^{\rm nlin}$) term, where two phonons are involved.
The electronic contribution to the decay rate of the $T$ phonon 
is then expected to be weak, and we will focus on the decay 
of the $L$ and $F$ modes. 

Let us start with the $L$ phonon with momentum $\hbar k$,
where the relevant lowest-order self-energy diagram due to  the first
term in $V_{e-ph}$ corresponds to the electron-hole bubble.  
Its imaginary part, responsible for the phonon decay rate, is 
$\propto [\delta(\omega-v_F k) - \delta(\omega+v_F k)]$, where $\omega=v_L k$.
(Note that we use a linearized dispersion for electrons.)
Since the Fermi velocity is $v_F\approx 10^6$~m/sec~$\gg v_L$, 
this condition can never be met, and 
only higher-order contributions can possibly lead to a phonon decay. 
This suggests that the decay rate of the $L$ phonon 
due to Eq.~(\ref{defpot}) is
very small, and probably negligible against the ph-ph mechanism.  

For the case of a $F$ phonon, the lowest-order contribution comes
from the second term in Eq.~(\ref{defpot}), leading to a ``fishbone''
diagram with two electron lines and a phonon line. The corresponding
imaginary-time self-energy is  given by
\begin{equation}
\tilde \Sigma_{F}(\tau,k) = -\frac{1}{\hbar \rho_M} \left ( \frac{3V_0}{2\pi R}
\right)^2 k^2 \sum_q (k-q)^2 {\cal G}(\tau,\omega_0) \Pi_e(\tau,q),
\end{equation}
with $\omega_0=\omega_\rho+ \hbar(k-q)^2/2m$ and
the 1D electron polarization function\cite{mora}
\[
\Pi_e( \tau,q) =  \frac{\omega_e}{2\pi v_F} \sum_{\xi=\pm} 
n(\xi\omega_e) e^{\xi\omega_e \tau},
\]
where $\omega_e = v_F |q|$.
Following the same steps as in Sec.~\ref{sec4}, we then find for the 
$T=0$ decay rate 
\[
\Gamma_F(k) \propto \sum_q \omega_e \delta(\omega_F(k)-\omega_0-\omega_e),
\]
which yields the energy conservation condition
\[
q^2 - 2qk + \sqrt{2}  \frac{v_F |q|}{v_L R} = 0.
\]
For $q\ne 0$,  this condition can only be met if $kR>v_F/(\sqrt{2} \ v_L)$,
i.e. only for short-wavelength phonons.
Thus, for the long-wavelength phonons of interest here, 
the energy mismatch between electron-hole pair excitations 
and the phonon modes implies that again only higher-order 
terms can possibly generate a finite decay rate.

The above discussion therefore suggests that e-ph couplings via the
deformation potential do not lead to significant decay rates of the
gapless $L$ and $F$ modes. Their decay should then be dominated by
the ph-ph interaction processes as described in Sec.~\ref{sec4}.

\section{Conclusions}\label{sec6}

In this paper, we have formulated and studied a general analytical
theory of phonon-phonon interactions for low-energy long-wavelength 
acoustic phonons in carbon nanotubes.
The continuum elasticity approach employed here reproduces 
the known dispersion relations of all gapless modes, 
including the flexural mode, $\omega_F(k)=\hbar k^2/2m$ with
``effective mass'' $m$.  We have then included
the most general cubic and quartic elastic nonlinearities allowed
by symmetry.
Remarkably, the relevant phonon-phonon
scattering processes giving the dominant contributions
to the decay rates are already found
from the geometric nonlinearities (i.e. using
the nonlinear strain tensor in a lowest-order expansion of the
elastic energy density), and for a quantitative
discussion of the decay rates, only the knowledge
of the well-known Lam{\'e} coefficients (or equivalently of the sound velocities) 
is necessary. 

We have provided  a complete classification of all possible three-phonon
processes involving gapless modes, along with the respective coupling
constants. At the level of four-phonon processes we have focussed on
the flexural modes, where a peculiarity is encountered, since the
four-phonon processes lead to a singular behavior of the finite-temperature
decay rate.  The physical reason for this singularity is the proliferation 
of phonons at finite temperature, implying a divergent 1D density of flexural modes.  
Interactions effectively regularize this divergence and lead to a finite density.
We have employed mean-field theory to quantitatively describe this
mechanism, and found a dynamical temperature-dependent gap 
$\hbar\omega_\rho$ for flexural phonons. While
this gap is in practice always below the thermal scale $k_B T$, it nevertheless
leads to important consequences and allows to compute a 
well-defined decay rate for flexural phonons in low-order perturbation
theory.  

Using this approach, we have determined the decay rate and the quality
factor for all long-wavelength gapless phonons in carbon
nanotubes.  We have also shown that electron-phonon interactions are
ineffective in relaxing those modes due to a mismatch in energy scales.
The reported quality factors ($Q$) are remarkably small,
especially for thin CNTs, pointing to important phonon
damping effects.  Phonon-phonon interactions in CNTs are therefore
significant and quite strong. 
Moreover, the values we have found are
rather close to the typical $Q$ reported in recent experiments.\cite{bachtold}
Our predictions represent intrinsic upper bounds for $Q$.  
Such upper bounds can be valuable in assessing the predictions 
of approximate theories, or when interpreting experimental data in 
terms of phonon damping.
We hope that our work will motivate further experimental and theoretical
studies along this line.

\acknowledgments

We thank I. Affleck for useful discussions.
This work was supported by the DFG SFB Transregio 12, by the ESF 
network INSTANS, and by the Humboldt foundation.

\appendix 
\section{Curvature effects}\label{appA}

In this appendix, we briefly discuss how to include curvature effects
in the nonlinear elastic energy.  To that end, we consider
the metric tensor for the cylindrical surface, 
whose components ($i,j=x,y=1,2$),
\[
g_{ij} =  \sum_{n=x,y,z} t_{n,i} t_{n,j},
\]
are expressed in terms of the ordinary scalar product
of tangent vectors 
\[
{\bf t}_i = \sum_{n} t_{n,i}({\bf r}) {\bf e}_n(x)
=\frac{\partial {\bf x}({\bf r})}{\partial x_i},
\]
with ${\bf x}$ given in Eq.~(\ref{xdef}). 
The nonlinear strain tensor then follows equivalently from $2u_{ij}=
g_{ij}-g_{ij}^{(0)}$,  with the metric tensor $g_{ij}^{(0)}=\delta_{ij}$ of 
the undeformed cylinder. 
The local unit normal vector is ${\bf N}= ({\bf t}_x\times
{\bf t}_y)/|{\bf t}_x\times {\bf t}_y|$, and the \textit{mean local curvature} $\Omega$
of the cylinder is defined as
\[
\Omega({\bf r}) = \frac12 \sum_{ij} b_{ij} g^{ij},
\]
where $g^{ij}$ is the inverse of $g_{ij}$, and the second fundamental
form of the surface is 
\[
b_{ij} =  \sum_{n=x,y,z}  N_n \partial_j  t_{n,i}.
\] 
To lowest order in the displacement fields $u_n$, the mean curvature  
is then given by\cite{suzuura}
\[
\Omega = \Omega_0 + \frac{u_z}{2R^2}  + \frac12 \left( \frac{\partial^2 u_z}
{\partial x^2} +\frac{\partial^2 u_z}{\partial y^2} \right),
\]
where $\Omega_0=-1/2R$ is the curvature of the undeformed cylinder.

Curvature leads to an additional energy cost due to hybridization effects. 
One can model this in a phenomenological way by adding an 
additional term 
\[
{\cal U}_{\rm curv}[u]= \kappa (\Omega-\Omega_0)^2
\]
to the elastic energy density, where $\kappa$ is a proportionality constant.
Such effects turn out to be small unless one deals with 
ultrathin CNTs, but they provide gaps to  
flexural modes with angular momentum $|\ell|>1$.
For $R\alt 1$~nm, these gaps are typically comparable in 
magnitude\cite{suzuura} to the breathing mode energy in Eq.~(\ref{breath}).

\section{Normal mode representation of the strain tensor}\label{appB}

In this appendix, we provide the explicit form of the strain
tensor expressed in terms of the normal mode displacement field operators
$u_J(k,\ell)$, see Eq.~(\ref{unewdef}). We keep all
$\ell=0$ modes ($L, T, B$), and the gapless flexural ($F$) modes with $\ell=\pm 1$.
For the linear part of the strain tensor, see Eq.~(\ref{ulin}), we obtain 
from Eq.~(\ref{displ}) the result
\begin{widetext}
\begin{eqnarray*}
u^{\rm lin}({\bf r})& =& \frac{1}{\sqrt{2\pi R} } \sum_k e^{iky} 
\Biggl [ \left( \begin{array}{cc}-i \nu k  &0 \\ 0& ik \end{array}\right ) 
u_L(k) + 
\left( \begin{array}{cc} 0 & ik/2 \\ ik/2 & 0 \end{array}\right ) u_T(k) +
 \left( \begin{array}{cc} 1/R & 0 \\ 0 & \nu k^2 R \end{array}\right ) u_B(k)  
\\ &&+ \frac{ik^2 R}{\sqrt{2}} \sum_{\ell = \pm} e^{i\ell x/R}
\left( \begin{array}{cc} \ell \nu & (1+\nu) k R \\
(1+\nu)k R & -\ell \end{array}\right ) u_{F}(k,\ell)
\Biggr ] ,
\end{eqnarray*}
while the nonlinear part (\ref{unlin}) reads
\begin{eqnarray*}
u^{\rm nlin}({\bf r})&=&\frac{1}{2\pi R}\sum_{k_1,k_2} e^{i(k_1+k_2)y}
\Biggl \{ 
\left( \begin{array}{cc}
 -\frac{\nu^2 }{2} k_1 k_2 & 0 \\ 0 & - \frac12 k_1 k_2
\end{array}\right ) u_L(k_1) u_L(k_2) \\ &+&
\left( \begin{array}{cc}
0 & \frac{\nu}{2} k_1(k_2-k_1) \\   \frac{\nu}{2} k_1 (k_2 -k_1)& 0
\end{array}\right ) u_L(k_1) u_T(k_2) 
\\ &+& \left( \begin{array}{cc}
 \frac{1}{2R^2} & 0 \\ 0 & - \frac12 k_1 k_2
\end{array}\right ) \left[ u_T(k_1) u_T(k_2) + u_B(k_1) u_B(k_2)\right ]\\
&+& \left( \begin{array}{cc}
0 & \frac{i}{2R} (k_1-k_2) \\   \frac{i}{2R} (k_1 -k_2)& 0
\end{array}\right ) u_T(k_1) u_B(k_2)  + 
\left( \begin{array}{cc}
-i\nu k_1/R &0  \\   0 & i\nu k_1^2 k_2 R
\end{array}\right ) u_L(k_1) u_B(k_2)\\ &
+&  \frac{1}{\sqrt{2}}\sum_{\ell_2=\pm} e^{ix\ell_2/R} \Biggl[
 \left( \begin{array}{cc}
\nu^2 \ell_2 k_1 k_2^2 R & (1+\nu)k_1 k_2/2 \\
(1+\nu)k_1k_2/2 & \ell_2 (k_2+\nu k_1) k_1 k_2 R 
\end{array}\right )  u_L(k_1) u_{F}(k_2,\ell_2) \\ & +& 
\left( \begin{array}{cc}
\nu k_2^2 & -\ell_2 k_2/2R   \\ -\ell_2 k_2/2R & - k_1 k_2
\end{array}\right ) u_T(k_1) u_{F}(k_2,\ell_2)  +
\left( \begin{array}{cc}
i\nu \ell_2 k_2^2 & i k_2/2R   \\ i k_2/2R & i\ell_2 k_1 k_2
\end{array}\right ) u_B(k_1) u_{F}(k_2,\ell_2) \Biggr ]
\\  & -& \frac{k_1 k_2}{4} \sum_{\ell_1\ell_2} e^{ix(\ell_1+\ell_2)/R}
\left( \begin{array}{cc}  1  & \frac{[ \ell_1 k_1 + \ell_2 k_2 +
 \nu ( \ell_1 + \ell_2)(k_1+k_2) ] R}{2} \\ 
 \frac{[ \ell_1 k_1 + \ell_2 k_2 + \nu (\ell_1+\ell_2)(k_1+k_2) ]R}{2}
 & (1-\ell_1\ell_2)  \end{array}\right ) 
\\  &&\times u_{F}(k_1,\ell_1) u_{F}(k_2,\ell_2)
\Biggr \}.
\end{eqnarray*}
\end{widetext}
For completeness, we also list the long-wavelength form of
the three-phonon amplitudes involving the breathing mode:
\begin{eqnarray*}
A_{BBB}&=& \frac{K+\mu+2\xi_1+2\xi_2} {2R^3}, \\
A_{BTT} &=& \frac{K+\mu} {2R^3}, \\
A_{LBB} &=&  \frac{i k_1}{R^2} [ -K+\mu+3(1-\nu)\xi_1 +(1-3\nu)\xi_2 ], \\
A_{LLB} &=& -\frac{k_1 k_2}{R} [ \nu K + 3(1-\nu^2)\xi_1+
(2\nu-1-3\nu^2)\xi_2]  ,\\
A_{BFF} &=& - \delta_{\ell_2,-\ell_3} \frac{(3K-\mu) k_2k_3}{4R} .
\end{eqnarray*}

\end{document}